\documentclass[10pt]{article}

\usepackage{indentfirst}
\usepackage{graphicx}
\usepackage{float}
\usepackage{caption}
\usepackage{multicol}
\usepackage{amsmath}
\usepackage{amssymb}
\usepackage{multirow}
\usepackage{color,soul}
\usepackage{mathtools}
\usepackage{algorithm}
\usepackage{algpseudocode}
\usepackage{lineno,hyperref}
\usepackage{apacite}
\usepackage{ragged2e}
\usepackage{booktabs}

\let\cite\shortcite

\usepackage{tabularx}
\usepackage{kantlipsum}
\newlength{\normalparindent}
\usepackage[T1]{fontenc}
\usepackage{newtxtext,newtxmath}

\usepackage{titlesec}
\newcolumntype{s}{>{\hsize=5\hsize}X}

\titleformat*{\section}{\fontsize{10}{14}\bfseries\filcenter}
\titlespacing{\section}{0pt}{\parskip}{-\parskip}

\renewcommand{\abstractname}{\vspace{-\baselineskip}}

\title{\fontsize{14}{14}\vspace{-0.5in} \bf
\textit{Privacy, Security, and Usability Tradeoffs of Telehealth from Practitioners' Perspectives}}
\author{}

\date{\vspace{-10ex}}

\begin{document}

\maketitle

\begin{abstract}
\centering
Faiza Tazi \textsuperscript{1}, Archana Nandakumar\textsuperscript{2}, Josiah Dykstra\textsuperscript{3}, Prashanth Rajivan\textsuperscript{2}, Sanchari Das\textsuperscript{1}\\
\textsuperscript{1} Department of Computer Science, University of Denver, Colorado, USA \break
\textsuperscript{2} Department of Industrial and Systems Engineering, University of Washington, Seattle, USA \break
\textsuperscript{3} Designer Security, LLC
\end{abstract}

\begin{abstract}
\centering
\begin{minipage}{6in}

\normalsize

The COVID-19 pandemic has significantly transformed the healthcare sector, with telehealth services being among the most prominent changes. The adoption of telehealth services, however, has raised new challenges, particularly in the areas of security and privacy.
To better comprehend the telehealth needs and concerns of medical professionals, particularly those in private practice, we conducted a study comprised of $20$ semi-structured interviews with telehealth practitioners in audiology and speech therapy. Our findings indicate that private telehealth practitioners encounter difficult choices when it comes to balancing security, privacy, usability, and accessibility, particularly while caring for vulnerable populations. Additionally, the study revealed that practitioners face challenges in ensuring HIPAA compliance due to inadequate resources and a lack of technological comprehension.  
Policymakers and healthcare providers should take proactive measures to address these challenges, including offering resources and training to ensure HIPAA compliance and enhancing technology infrastructure to support secure and accessible telehealth.
\end{minipage}
\end{abstract}{\vspace{0ex}}

\begin{multicols}{2}

\vspace{5mm}
\section*{INTRODUCTION}
\indent{}

The COVID-19 outbreak, declared a global pandemic by the World Health Organization in 2020, had a profound impact on healthcare systems and patients worldwide. In response, telehealth services experienced an unprecedented surge with a record growth of 64.3\% in 2020~\cite{ahip2020}. 
The use of telehealth technologies however raises privacy and security concerns given the sensitive health information they transmit and store. This information is vulnerable to cyberattacks, theft, and data breaches which can lead to unauthorized access, manipulation, or destruction of telehealth systems and data. To mitigate these risks, healthcare providers and technology companies must comply with legal and regulatory requirements such as the Health Insurance Portability and Accountability Act (HIPAA) and the General Data Protection Regulation (GDPR), which may require the implementation of robust security measures including encryption, multi-factor authentication, and security updates~\cite{albanese2021medicare,ong2020inpatient}. 

Medical professionals play critical roles in healthcare, yet research on their privacy and security perspectives is limited, particularly in private practice telehealth~\cite{tazi2022sok}. Audiology and speech-language pathology represent two important components of allied healthcare where telehealth has the potential to revolutionize service delivery. Ensuring the privacy and security of patients' sensitive health information is crucial for the success of any telehealth service. Unfortunately, the severe lack of research in this area is concerning. To address this gap, we conducted $20$ semi-structured individual interviews with audiologists and speech-language pathologists (SLPs) to gain insights into how telehealth is used to provide care and therapy for patients. 
This paper aims to (1) identify privacy and security concerns of audiologists and SLPs in the telehealth context, (2) understand the perceptions of audiologists and SLPs towards HIPAA compliance during telehealth sessions, and (3) examine providers' experience and sentiment towards telehealth.

\section*{METHODS}

This study investigates the current understanding and perception of privacy and security among audiologists and speech-language pathologists (SLP) in private  practice. Through this research, we focus on the use of telehealth technologies in these practices. To gather data for the study, the research team designed and conducted semi-structured interviews with $20$ audiologists and SLPs who are practicing in the United States. For flexibility to the interviews and given the research theme, we conducted the interviews online using Zoom between August 2022 and January 2023. Through this study, we aim to answer the following research questions:
\begin{itemize}
\item \textit{RQ1: How do audiologists and speech-language pathologists in private practice utilize telehealth technologies in their clinical work? What are the perceived benefits and challenges of using telehealth technologies in the practices of audiologists and speech-language pathologists?}
\item \textit{RQ2: What is the current understanding and perception of privacy and security among audiologists and speech-language pathologists practicing in private practice settings? How do audiologists and speech-language pathologists ensure the privacy and security of patient data when using telehealth technologies in their practices?}
\end{itemize}

\subsection*{Recruitment}
We began the recruitment process for the study after obtaining the necessary Institutional Review Board (IRB) approval from relevant institutions. Given the nature of the study, we worked closely with the IRB to ensure we followed the standard ethical practices. We implemented a multi-pronged approach, interviewing $20$ participants until we reached the qualitative data collection standards. Initially, we advertised the study in various social media groups specific to speech-language pathology and audiology professionals. We also circulated the recruitment emails across various mailing lists, which the researchers knew focused on our targeted participant pool. We also implemented snowball sampling to identify potential participants. However, there needed to be more outreach than these methods since we focused on participants who used telehealth technologies for a considerable time (at least two years). 

To overcome this recruitment challenge, the research team contacted the two professional organizations for audiologists and speech-language pathologists, the American Speech-Language-Hearing Association (ASHA) and the Academy of Doctors of Audiology (ADA). These organizations were instrumental in recruiting participants for the study, as they sent emails to their affiliated members on behalf of the research team. By the end of the recruitment phase, we obtained a diverse sample of $20$ interview participants from the United States. The sample comprised $10$ audiologists and $10$ speech-language pathologists with varying years of medical expertise and experience with telehealth technologies. This diversity in expertise and experience provided valuable insights and perspectives on using telehealth technologies in private practices. We provide the demographic details of our participants in Table~\ref{tab:demographics}.

\begin{table*}[t]
\renewcommand{\arraystretch}{1} 
\selectfont\centering
{\small
\begin{tabular}{cccccc}
\toprule
{\textbf{ID}}
 & {\textbf{Role}} 
 & {\textbf{Gender}} 
 & {\textbf{Work Exp.}}
& {\textbf{Telehealth Exp.}}
& {\textbf{Platform(s)}}\\
\midrule

P1  & AuD    & M & 6-10 & 1-5  & Tuned + Zoom + Google Meet\\
P2  & AuD    & M & 11+  & 6-10 & Tuned + Zoom + Google Meet \\
P3  & AuD    & M & 11+  & 1-5  & Blueprint\\
P4  & AuD    & F & 1-5  & 1-5  & Zoom\\
P5  & AuD    & F & 6-10 & 1-5  & CounselEAR\\
P6  & AuD    & F & 6-10 & 1-5  & CounselEAR\\
P7  & AuD    & F & 11+  & 11+  & CounselEAR\\
P8  & AuD    & F & 11+  & 1-5  & Zoom\\
P9  & AuD    & F & 11+  & 1-5  & Modmed+Athena\\
P10 & AuD    & F & 6-10 & 6-10 & Zoom\\
P11 & SLP    & F & 1-5  & 1-5  & Google Meet\\
P12 & SLP    & F & 11+  & 1-5  & Zoom\\
P13 & SLP    & F & 11+  & 6-10 & Theraplatform\\
P14 & SLP    & F & 11+  & 1-5  & SimplePractice\\
P15 & SLP    & F & 11+  & 1-5  & Zoom\\
P16 & SLP    & F & 6-10 & 1-5  & Theraplatform\\
P17 & SLP    & F & 6-10 & 1-5  & Zoom or Google Meet\\
P18 & SLP    & F & 11+  & 1-5  & Zoom + Google Meet + doxy.me + Blink Session\\
P19 & SLP    & F & 11+  & 1-5  & Televote (proprietary platform)\\
P20 & SLP    & F & 6-10 & 1-5  & Zoom\\
\bottomrule
\end{tabular}
}
\caption{Participant demographics. 
}
\label{tab:demographics}
\vspace{-3ex}
\end{table*}


\subsection*{Interviews}
After distributing recruitment materials, potential participants reached out to the research team via the provided email. We confirmed that $21$ of the $104$ people who contacted us were indeed audiologists or SLPs and we scheduled online interviews with all $21$ of them. One participant had to cancel their interview. All of the participants appreciated the flexibility of the online interview. We conducted the interviews between August 2022 to January 2023, and each interview lasted an average of $46$ minutes ($min=32\ minutes$, $max=90\ minutes$). Each participant received an electronic gift card of USD $\$50$ for their participation in the study. We designed the questions to be open-ended, encouraging participants to provide detailed and in-depth discussions. The semi-structured interviews began with a brief explanation of the study and the obtaining of verbal consent, including permission for a Zoom recording and general consent to participate in the study. All but one participant agreed to have their interview recorded on Zoom. 

\subsection*{Data Analysis}
Our data analysis was broken down into several stages. The first included a verbatim transcription of the interviews. Because only one participant refused to be recorded, the interview administrator took notes of the responses provided while the remaining participants were audio recorded and transcribed. These notes and transcriptions were then anonymized. The second phase of our study consisted of a qualitative analysis approach using thematic analysis. One research team member conducted inductive coding and generated an initial codebook based on the research questions and the initial coding of two transcripts. A separate team member conducted independent coding on these two transcriptions using previously generated codes to ensure inter-rater reliability (IRR). These two team members independently coded the remaining $17$ transcripts and the one interview note. The coders then discussed changes or additions to the codebook as they arose to refine the codebook and group codes into themes. Once the coding was complete, data were organized into tables and figures to represent the findings visually. Additionally, we used quotes from the transcripts to support the findings and give voice to the participants. 

\section*{FINDINGS AND DISCUSSION}

Across both specialties, our participants reported a remarkably similar workflow. A telehealth session will typically begin a few minutes before the appointment time. This additional time allows the providers to review the patient's history, prepare their materials, and ensure that their software is operational. When the patient joins the session, the disparities between participants begin to manifest. Some participants agreed that they will first confirm the patient's identity before beginning a session, while others like P9 said they do not: 

\begin{quote}
\lq\lq Most of the people that I'm doing telehealth with\ldots I know these patients\ldots I know their face.\rq\rq~
\end{quote}

The services provided during these telehealth sessions varied by specialty, with SLPs providing services ranging from initial evaluations to therapy treatments. Among our participants, telehealth services for audiologists included follow-up appointments, consultations, and hearing aid programming. Moreover, our participants reported using a variety of telehealth platforms: Athena, Blink session, Blueprint, CounselEar, doxy.me, Google Meet, Modmed, Televote, Theraplatform, Tuned, SimplePractice, and Zoom. 
Qualitative analysis produced four overarching themes described below: (1) technological disparity, (2) flexibility with telehealth, (3) HIPAA confusion, and (4) training.

\subsection*{Technological Disparity}

\subsubsection*{Privacy and security concerns}

%
%
Some of our participants were aware of the threats they face when delivering telehealth services or simply utilizing an Electronic Medical Record (EMR) system; as a result, these participants implemented security measures to safeguard both the practice and the patients, despite knowing that these safeguards can never be 100\% failsafe. As P5 noted:
\vspace{-2ex}
\begin{quote}
    \lq\lq We don't have patient data just randomly saved places\ldots we're doing the best that we can\ldots It's why I make sure when we're doing those telehealth visits we are in our EMR and not on Zoom or not on Teams or on FaceTime
    \rq\rq~
    \end{quote}

When asked about the security measures put in place to protect telehealth sessions, seven participants mentioned enabling waiting rooms. This is especially critical for ensuring that patients are not unintentionally or purposely exposed to the personal health information of other patients. In addition, only two passwords plus a password manager. Likewise, only two people mentioned using a secure internet connection. Five participants affirmed that they do not record any telehealth sessions, with one stating that even though they do not record their sessions with patients they still include it in their consent forms \lq\lq in case [the provider] make a mistake\rq\rq.

We also identify participants who are cognizant of the importance of privacy and security but are unconcerned about it, either because they have an IT department managing security or because they chose a secure software provider they trust. P3's note on Blueprint is an example of this trust:
\vspace{-1ex}
\begin{quote}
    \lq\lq 
    So Blueprint has their own server in-house they show us their security they haven't been hacked or anything like that. It's a very well protected, well funded, high quality secure service, and again that's why we chose them. There's a lot of choices out there but that's why we chose them as our office management software.\rq\rq~
\end{quote}
Furthermore, some participants were unconcerned about the privacy and security of their telehealth sessions, justifying their apathy by claiming that the confidential healthcare information disclosed during telehealth meetings is restricted and of little utility to bad actors. 
As P11 put it:
\vspace{-1ex}
\begin{quote}
    \lq\lq Of all the things on this earth for them to want to hack I don't think that our therapy appointments are going to be the biggest priority and if they do they'll just come in and see us playing games online and practicing speech sounds.\rq\rq 
\end{quote}

\subsubsection*{Usability}
When asked about the reason behind their software choices, eight participants mentioned that \lq\lq ease of use\rq\rq~was an essential consideration for both their patients and themselves. Subsequently, participants confirmed the discontinuation of telehealth software use that was difficult for their patients to use, navigate, or comprehend. This is especially true for participants who serve communities in difficult socioeconomic situations and older populations who have difficulty using telehealth, as P15 stated:   
\begin{quote}\lq\lq
    [Zoom] wasn't as easy as just click on this button and you can enter my teletherapy space\ldots It was just too many steps for the population I was working with, and so it ended up just being very frustrating for them. So I ended up using Google Meet mostly because it was easier for the families to join the meeting\ldots
    \rq\rq~\end{quote}
These participants revealed that usability is critical to ensuring continuity of care, especially during COVID-19, as patients may be less likely to use a telehealth platform if it is difficult to use.

\subsubsection*{Cost}
When deciding on telehealth platforms, cost is an essential consideration, particularly for providers in small private practice facilities ($4$ full-time employees or less) or solo practitioners. These providers must manage their budgets efficiently while still providing high-quality care to patients. Some telehealth platforms can be costly, and when the cost is prohibitively expensive, some of our participants decided to switch platforms, occasionally sacrificing functionality to reduce costs. Although none of our participants compromised HIPAA compliance, as P11 stated:
\begin{quote}\lq\lq
    \ldots We needed [an] alternative. So what are the cost effective alternatives out there? Google Meet. Can we make it HIPAA compliant?\ldots and once we have the [Business Associate Agreement] in place was when I felt most comfortable using Google [Meet]\rq\rq~\end{quote}

%
%
\subsubsection*{Resource Availability}
Despite the availability of online resources to help providers select and use telehealth, it can be challenging for them to identify the most relevant ones for their needs and situations, according to some of our participants. For instance, some needed help finding guidance for maintaining HIPAA compliance when offering telehealth. Some of our participants also struggled with finding a platform suitable for offering telehealth service, as P1 mentions: 
    \begin{quote}\lq\lq To be honest I'm not entirely sure what else is available [other than Google Meet]\ldots 
    \rq\rq\end{quote}
This could be because telehealth is a relatively new field, so fewer established resources may be available than there are for conventional healthcare services. Furthermore, the telehealth landscape quickly evolves with new technologies and regulations, making it challenging to keep up with the most recent information. This has led some participants to choose a specific platform for the resources and information it offered them.
%
%
%
This is primarily because readily available information, especially when it is from a HIPAA-compliant source, will help providers make informed decisions while improving productivity and saving time and resources, as P9 mentions: 
    \begin{quote}\lq\lq I don't have time and my job is not to do cyber security; my job is to take care of patients\rq\rq~\end{quote}

\subsection*{Flexibility with Telehealth}

\subsubsection*{COVID Considerations}

When asked about the reasons for providing telehealth, our participants stated that telehealth enabled patients to access healthcare services without physically visiting a clinic or hospital, which was especially important at a time when social isolation was required to slow the spread of COVID-19. Furthermore, by eliminating the need for in-person visits, telehealth reduces the risk of infection for patients and healthcare workers. This assisted in protecting our participants as well as their vulnerable patients from COVID-19 exposure. Notably, many participants emphasized the importance of telehealth in enabling patients to continue receiving care from their regular healthcare providers, thereby ensuring care continuity. This was especially essential for patients with chronic illnesses who needed continuous care and monitoring; P14 confirms this in their statement:
    \begin{quote}\lq\lq We had a lot of medically fragile children\ldots we got together and did not want our patients to go without services\ldots that's when we started learning about teletherapy.\rq\rq~\end{quote}

\subsubsection*{Physical Harm}
Telehealth enabled healthcare providers to provide care virtually, reducing the need for in-person visits and reducing the risk of physical harm or violence. This was especially essential for healthcare providers who deal with patients who exhibit aggressive behavior or work in environments where they feel unsafe. This is the case of two of our interviewees who agreed that physical safety is one of the reasons they choose to provide telehealth services. 
These participants also claimed that they no longer feel safe delivering in-person care in ``high-risk schools'' because the environment can impact therapy quality. As P17 puts it: 
\begin{quote}\lq\lq I kind of worked in a school that had a lot of violence. The school I was in\ldots it was definitely something I had to be monitoring for and very aware of all the time. So that made me feel more secure doing teletherapy, just like the safety of my own physical safety was not a concern which was like a very normal part of my day\ldots that can all distract from therapy goals and take time out of your day \rq\rq~\end{quote}

\subsection*{HIPAA}

\subsubsection*{HIPAA Confusion}
HIPAA is a complex federal law with provisions and regulations that can be challenging for non-legal experts to understand and implement. This may lead to misconceptions or misunderstandings concerning HIPAA requirements which can further confuse healthcare providers and make understanding and complying with the legislation problematic. Some participants were confused about HIPAA compliance when using platforms such as Google Meet and Zoom.
%
%
Furthermore, the regulatory framework may have yet to keep up with the rate of change in the telehealth sector, mainly after COVID-19 has spurred the adoption of the technology and given rise to new use cases. As a result, there may be some uncertainty and ambiguity around how HIPAA regulations apply to telehealth technologies. As stated by P9: 
    \begin{quote}\lq\lq HIPAA, but not with telehealth, I think that's still a gray area. I think there's not a real defined protocol for telehealth as it pertains to HIPAA. We're told what HIPAA is and how we need to define it and what is considered a breach, but when it comes to these different areas\ldots I mean telehealth is kind of underneath that umbrella. I mean, do I really know if something's HIPAA compliant? I don't think the administrators even know.\rq\rq~\end{quote}

\subsubsection*{HIPAA Violations}
Some of our participants acknowledged breaking HIPAA laws. Indeed, P16 stated that they could not keep a telehealth talk from being overheard because they did not have a soundproof space in their home. This could be considered a violation since HIPAA requires covered entities to protect the privacy and security of PHI, which includes information about an individual's health status or healthcare services obtained:
\begin{quote}\lq\lq
     During COVID it was hard because I'm working from home and my husband was here and he could very easily hear my sessions and would hear me use patients names and things like that and we didn't talk about the patients\ldots but he knew what my patients names were because he was in the house so I feel like that kind of thing was understood as something that is unavoidable and inevitable when everyone is working from home and on lockdown.
     \rq\rq~\end{quote}

Furthermore, some healthcare workers may be unfamiliar with the security features and requirements of the telehealth platform they are using, potentially resulting in unintentional HIPAA violations. This is the case of some interviewees who described utilizing FaceTime for telehealth or saw other clinicians using it in a telehealth session. While the U.S. government temporarily relaxed HIPAA enforcement for the use of FaceTime and other non-public telehealth technologies during the COVID-19 public health emergency, enforcement is scheduled to return in 2023. Finally, HIPAA violations may be due to a lack of training if healthcare workers are not sufficiently knowledgeable of HIPAA regulations, particularly with regards to telehealth.

\subsection*{Training}
According to the law, and confirmed by our study participants, healthcare personnel who work with PHI must complete periodic HIPAA training. HIPAA training is intended to inform healthcare professionals of the law's obligations and show them how to manage PHI per its guidelines. Several participants opt to enroll in continuing education programs to fulfill the required training. HIPAA does not, however, mandate any telemedicine-specific training for telehealth providers. Nevertheless, telehealth education is crucial because, unlike conventional in-person healthcare, telehealth calls for unique skills and knowledge. Healthcare professionals should thus undergo telehealth training, given that it can help them provide high-quality care to patients who cannot access in-person care. As a result, more patients will be able to obtain care, particularly those who reside in isolated or underserved locations or find it challenging to leave their homes. As a result, only four participants indicated that their states' laws required them to complete telehealth training, as reported by P20: \begin{quote}\lq\lq In Massachusetts they technically require [telehealth] training for all speech pathologists.\rq\rq~\end{quote}

\section*{FUTURE WORK AND LIMITATIONS}
The study provides valuable insights into the privacy and security perspective of audiologists and SLPs regarding telehealth. However, limitations exist due to the qualitative nature and small sample size of 20 participants. To enhance the generalizability of findings, future studies should include a broader range of healthcare professionals from various geographical locations. This would provide a more comprehensive understanding of privacy and security concerns in telehealth. Additionally, future research could explore the use of telehealth technologies in other medical specialties and how privacy and security concerns differ. Finally, examining the impact of privacy policy regulations changing landscapes could further inform the development of privacy-preserving telehealth technologies.

\section*{CONCLUSION}
Protecting and securing medical data while maintaining privacy is essential in all aspects of healthcare, particularly when using third-party telehealth services that lie beyond the scope of medical institutions' policy management. Therefore, we conducted $20$ interviews with audiologists and SLPs from private practices in the United States to understand their perspective on telehealth, including their usage, technology selection, and privacy and security considerations. Our data analysis revealed valuable insights into the telehealth landscape from the audiologists' and SLPs' viewpoint. A key finding was that the providers are concerned and motivated to ensure the security and privacy of their telehealth patients but do not have access to necessary resources to enforce them with little uncertainty. The main barriers appear to be cost, training, usability and workflow in these telehealth systems. Privacy and security were also emphasized as crucial factors, with participants often relying on the services they use for data protection. Finally, our study highlights the need for continued education and training in telehealth to address challenges such as navigating HIPAA regulations and selecting appropriate telehealth platforms. Further research is necessary to identify barriers to telehealth adoption and optimize it to meet audiologists' and SLPs' needs. This information can aid in developing targeted training and education programs to support effective telehealth implementation in private practices.

\section*{ACKNOWLEDGEMENT}
We would like to thank our participants for their time and input and acknwoledge the Inclusive Security and Privacy focused Innovative Research in Information Technology (InSPIRIT) Laboratory at the University of Denver. This research has been funded by CISCO Research Award. Any opinions, findings, conclusions, or recommendations expressed in this material are solely those of the authors and not of the organization or the funding agency.
\indent{}
\bibliographystyle{apacite}
\renewcommand{\refname}{\fontsize{10}{14} REFERENCES}
\titlespacing\section{0pt}{0pt}{2ex}
\renewcommand\bibliographytypesize{\footnotesize}
\\
\bibliography{HFES/HFES}
\addtolength{\textheight}{-11cm}   

\end{multicols}

\end{document}